\documentclass[fleqn,usenatbib]{mnras}

\usepackage{newtxtext,newtxmath}
\usepackage[T1]{fontenc}

\DeclareRobustCommand{\VAN}[3]{#2}
\let\VANthebibliography\thebibliography
\def\thebibliography{\DeclareRobustCommand{\VAN}[3]{##3}\VANthebibliography}

\usepackage{graphicx}	% Including figure files
\usepackage{amsmath}	% Advanced maths commands
\title[Interaction of CMB photons in galaxy halos]{The cosmic shallows
I: interaction of CMB photons in extended galaxy halos}

\author[H. E. Luparello et al.]{
Heliana E. Luparello,$^{1}$\thanks{E-mail: heliana.luparello@mi.unc.edu.ar (HEL)}
Ezequiel F. Boero,$^{1,2}$ 
Marcelo Lares,$^{1,2}$ 
Ariel G. S\'anchez$^{3,4}$
\newauthor
and Diego Garcia Lambas$^{1,2}$
\\
$^{1}$Instituto de Astronom\'{\i}a Te\'{o}rica y Experimental,
CONICET-UNC, Córdoba, Argentina\\
$^{2}$Observatorio Astronómico de Córdoba, UNC, Córdoba, Argentina\\
$^{3}$Max-Planck-Institut f\"ur extraterrestrische Physik, Giessenbachstr. 1, 85748 Garching, Germany\\
$^{4}$Universit\"as-Sternwarte M\"uchen, Fakult\"at f\"ur Physik, Ludwig- Maximilians-Universit\"at M\"unchen, Scheinerstrasse 1, 81679 M\"uchen, Germany
}

\date{Accepted XXX. Received YYY; in original form ZZZ}

\pubyear{2021}

\begin{document}
\label{firstpage}
\pagerange{\pageref{firstpage}--\pageref{lastpage}}
\maketitle

\begin{abstract}
We report and analyse the presence of foregrounds in the cosmic
microwave background (CMB) radiation associated to extended galactic halos.
Using the cross correlation of \textit{Planck} and \textit{WMAP} maps and the 2MRS galaxy catalogue, we find that the mean temperature radial profiles around nearby galaxies at $cz\le 4500~\rm{km~s^{-1}}$ show a
statistically significant systematic decrease of $\sim 15~\mu \rm{K}$ extending up to several galaxy radii.
This effect strongly depends on the galaxy morphological type at scales within 
several tens of times the galaxy size,
becoming nearly independent of galaxy morphology at larger scales.
The effect is significantly stronger for the more extended galaxies, with galaxy clustering having a large impact on the results. 
Our findings
indicate the presence of statistically relevant foregrounds in the CMB maps that should be considered in detailed cosmological studies.
Besides, we argue that these can be used to explore the intergalactic medium surrounding bright late-type galaxies and allow for diverse astrophysical analyses.
\end{abstract}


%%%%%%%%%%%%%%%%%%%%%%%%%%%%%%%%%%%%%%%%%%%%%%%%%%

%%%%%%%%%%%%%%%%% BODY OF PAPER %%%%%%%%%%%%%%%%%%

\section{Introduction}

The prediction of the cosmic microwave background (CMB) by Alpher and
Gamow \citep{Gamow1946, Alpher1948} as a result of
their findings in connection with the origin and formation of light
elements in the early Universe is nowadays one of the central pillars of 
modern cosmological models \citep[e.g., ][]{Dodelson2003}.
Since the decade of 1960, when its discovery and interpretation as a
signal of cosmological origin was achieved
\citep{Penzias1965, Dicke1965}, this primordial radiation
has provided a wealth of knowledge on the early stages of the
Universe. 
The background radiation is highly isotropic, with 
small variations of fundamental relevance to the 
current scenario on the evolution of large-scale
structures \citep{Peebles1980}, and it is crucial in the
derivation of cosmological parameters.
At late times, the local Universe is highly non-homogeneous and the 
anisotropies in the CMB radiation can be useful for the study of the foreground matter fields. 
For example, \citet{DePaolis:2019hnq} use the CMB spectrum to analyse
the discrepancies between local measurements of 
the abundance of
baryons and observations based on the large-scale
structure of the Universe.

Understanding foreground effects in the CMB spectrum is fundamental given its relevance for accurate
cosmological parameter determinations as well as insights on a detailed understanding of the 
astrophysics of the intergalactic medium.
It is useful to categorise foregrounds according to their impact on different angular
scales on the sky as well as their detection as diffuse or compact emission. At large angular 
scales, the main  contamination of the CMB radiation is due to extended emission
from our galaxy and can be identified with several well known processes such as synchrotron radiation, 
or thermal bremsstrahlung.
There are several works that consider
different issues related to possible foregrounds. For instance,
\citet{Afshordi2004} analysed cross correlation between the
2MASS survey \citep{Jarret:2000} and \textit{WMAP} data \citep{Bennet:2003b}.  
In their analysis, the authors study the presence of the integrated Sachs-Wolfe effect \citep[ISW, see
also][]{Rassat2006, Francis2009a,Francis2009b} and hot gas (thermal
Sunyaev-Zeldovich effect; thermal SZ), as well as microwave point
sources.
The authors consider the 2MASS catalogue in bins of
K-band magnitude that define mean redshift bins ranging from $15000$ to $30000~\rm{km~s^{-1}}$.
Similarly, other authors have focused on SZ signature of baryons
\citep{hernandez_thermal_2006} or the raise of non gaussian effects
due to SZ-induced features \citep{Cao2006}.
Also, some of the large-scale CMB anomalies have been examined and removed after
subtraction of the integrated Sachs-Wolfe effect by
\citet{Rassat2013}.
%
%XXX
Using data from the \textit{Planck} satellite, \citet{Planck_XI} determined the scaling 
relation between the Sunyaev-Zeldovich effect and the stellar masses of local bright 
galaxies from the NYU-VAGC \citep{NYUVAGC}. Most of these galaxies are 
the central galaxies of their dark matter halos, and the signal is present even for 
low stellar masses, indicating the presence of hot and diffuse gas in those systems.
More recently, \citet{Makiya2018} have explored the thermal~SZ
effect associated with 2MRS galaxies at redshift $z\simeq 0.03$.
They analysed the auto and cross correlations between the power spectra of the thermal 
SZ effect and the galaxy number density. 
These measurements, along with CMB data, provide constraints for the mass bias indicating that the estimated \textit{Planck} cluster masses  should be 
35 per cent lower than the true masses.
Through the correlations they also study the electron pressure profile,
providing important clues at
understanding gas physics in the local Universe.
%
%XXX

%
In this work, we focus on the cross correlation of \textit{Planck} data  \citep[i.e. CMB 
temperature and polarisation amplitude,][]{planck2018IV} with the location and properties 
of galaxies from the 2MASS galaxy redshift survey \citep[2MRS,][]{Huchra2012_2MRS}. 
The CMB data, as well as the galaxy samples are described in Section~\ref{cats}.
A brief description of our methodology of analysis is given in Section \ref{methods}.
In Section~\ref{profiles} we explore the CMB temperature profiles around the 
galaxies in our sample for a wide range of scales and morphological types, and assess 
the influence of different galaxy environments.
To further explore the nature of the signal, in Section \ref{models} we propose a simple 
model that recovers the main traits of the observational data.
In Section \ref{polarisation} we explore the CMB polarisation flux around the galaxies.
In Section \ref{discussion} we summarise our main results and discuss their possible implications.
In Appendix \ref{app_z}, we show the results for different redshift ranges.
We assess the reliability of our results using the nine-year observations of the \textit{WMAP}
satellite \citep{2013ApJS..208...20B}.
In Appendix \ref{app_wmap}, we show the mean temperature profiles using data from \textit{Planck} and
\textit{WMAP}. 
With these tests, we exclude the
possibility that our findings are subject to instrumental effects.
In Appendix \ref{app_sz} we analyse the influence of the thermal Sunyaev-Zeldovich effect.
\section{Data}\label{cats}
Our joint analysis of galaxy positions and CMB temperature and polarisation maps is based
on the data products released in 2018 by the \textit{Planck}
collaboration hereafter PR3 \citep{Aghanim2020b,Aghanim2020c,Aghanim2020a}
together with the 2MRS.
We have also addressed these analyses using the final \textit{WMAP} temperature maps \citep{Bennett2013}.

\subsection{CMB maps}\label{subsec:CMBmaps}

There are several maps with implementation of  foreground removal algorithms.
We use the SMICA CMB temperature and polarisation maps of \textit{Planck} PR3 \citep{planck2018IV}.
These maps are expected to have small foreground residuals across a wide range of angular scales \citep{PlanckXII_2013, PlanckIX_2015}.
The cleaned maps from PR3 have also a better control of 
systematic effects. 
The main improvements with respect to previous releases are in polarization, as described in
\citet{planck2018IV}.

In our analysis, we use the maps produced with the data 
from the full mission period containing information from the nine frequency channels.
%XXX
%For a proper analysis, we also used the provided common confidence masks for temperature and polarisation.
For a proper analysis, we also used the provided common confidence masks for temperature and polarisation.
%XXX
%
%The Planck Collaboration released four different CMB cleaned maps 
%corresponding to four different pipelines to deal with foregrounds.  
%We have performed our analysis using the SMICA dataset.
%
We used a map resolution given by $\rm{N}_{\rm{side}} = 2048$, according to the
HEALPix \footnote{\url{https://healpix.sourceforge.io}} software applied for the tesselation of the 
celestial sphere.
The area of each pixel in this setting is about $\Omega_{\rm{p}} = 1.062 \times 10^4$~$\text{arcsec}^2$.
Data from \textit{WMAP} \citep{2013ApJS..208...20B} was additionally used to reinforce our analysis as an 
independent consistency check of the results.
An account of our results with this dataset is given in Appendix~\ref{app_wmap}.
%
%The details of this analysis are given in the Appendix~\ref{app_wmap}.

\subsection{Galaxy catalogue}

We explored the properties of the CMB radiation around galaxy samples 
taken from the 2MRS.
This catalogue contains approximately 44\,600 galaxies with  magnitude in the K-band $K_s
\leq 11.75$~mag and galactic latitude $|b| \geq 5^{\circ}$ distributed across the
whole sky with an area coverage of 91 per cent. 
More than 11\,000 galaxies have spectroscopic observations, and
a subsample of 20\,860 galaxies ($K_s\leq11.25$~mag and $|b|\geq10^{o}$) have morphological type 
classification  according to the
modified Hubble sequence \citep{devaucouleurs_revised_1963,Huchra2012_2MRS}.

The 2MRS catalogue has a wide sky coverage and completeness, while 
it provides essential astrophysical data such as redshift, morphology and optical radius.
The angular size of the galaxies is characterised by the isophotal radius enclosing the 
total K-band magnitude of the galaxy ($r_{\rm{ext}}$, as defined in the 2MRS database).
These properties make this catalogue an ideal dataset to study the possible presence of 
foreground absorption associated to relatively close sources.
\begin{table*}
   \centering
   \caption{Number of galaxies as a function of the redshift range for the samples used in this work. Our main samples are in the redshift range $300~\rm{km~s^{-1}}<cz<4500~\rm{km~s^{-1}}$, and the supplementary samples in the range $300~\rm{km~s^{-1}}<cz<12000~\rm{km~s^{-1}}$ are analysed in the Appendix \ref{app_z}.}
\begin{tabular}{ |p{5cm}||p{2cm}|p{2cm}|p{2cm}| p{2cm}|p{2cm}| }
\hline
Redshift range  & E & S & Sa & Sb & Sb+Sc+Sd    \\
\hline
$300~{\rm km~s^{-1}} < cz <4500~{\rm km~s^{-1}}$  & 524   & 3589  & 1100 & 1147 & 2489 \\
$300~{\rm km~s^{-1}}< cz <12000~\rm{km~s^{-1}}$  & 3445  & 12\,556 & 4\,520 & 4\,388 & 8036 \\
\hline
\end{tabular}
\label{tab:T01}
\end{table*}

\begin{figure}
	\includegraphics[width=\columnwidth]{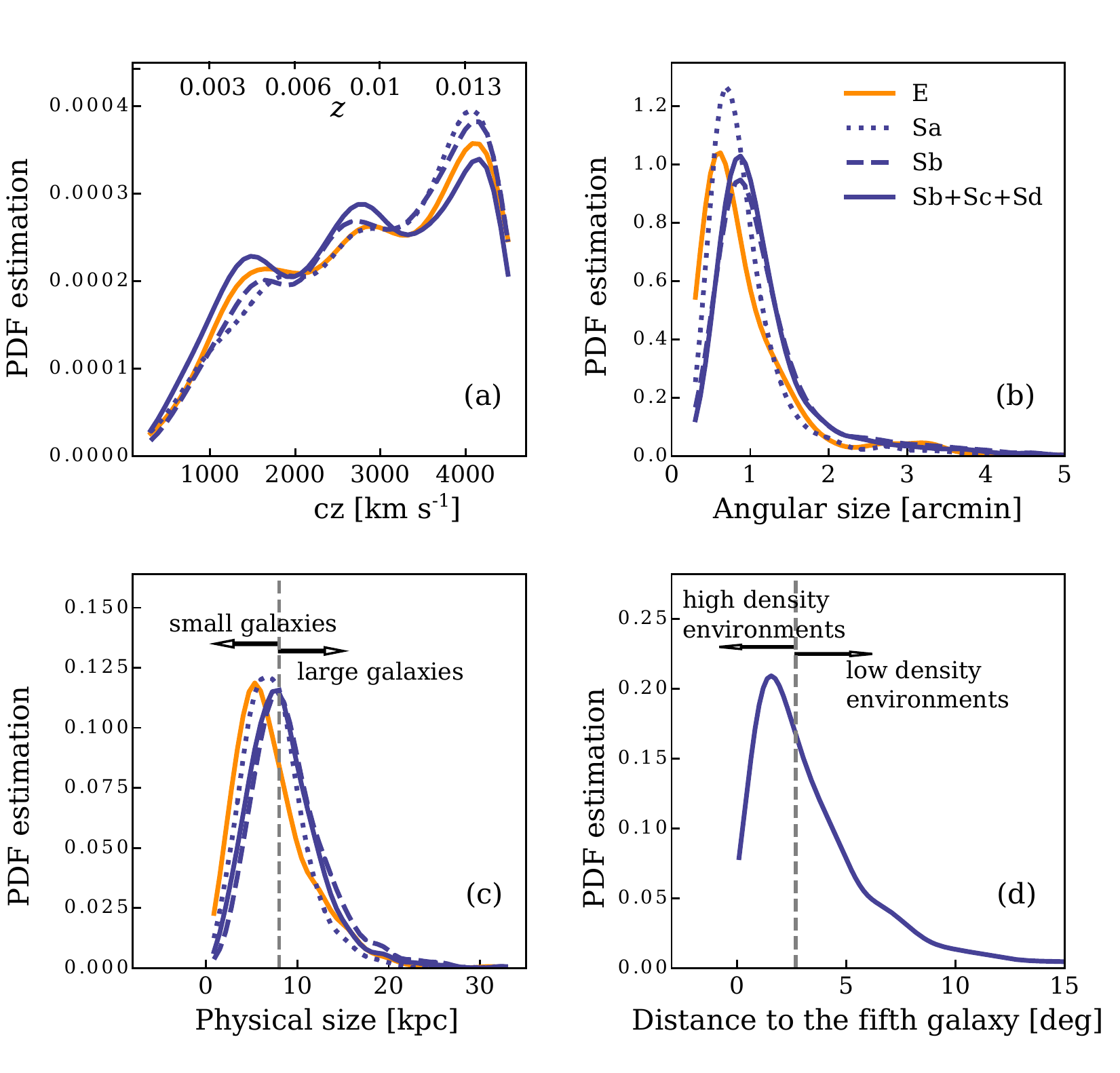}

    \caption{Distribution of redshift (panel (a), in $\rm{km~s^{-1}}$), angular galaxy size (panel (b), in arcmin), physical size (panel (c), in $\rm{Kpc}$) and local density parameter (panel (d)), estimated as the distance to the fifth nearest neighbour galaxy.
    In all panels the orange lines correspond to elliptical galaxies, and the blue lines correspond to the different subtypes of Spiral galaxies. The dashed vertical lines in the bottom panels show the median values of the distributions for the Sb~+~Sc~+~Sd galaxies, used to split this subsample according to the galaxy sizes and its environment.
    }
    \label{fig:F01}
\end{figure}

We consider galaxy samples defined by two distance limits within \mbox{$cz= 12\,000~\rm{km~s^{-1}}$}.
Table~\ref{tab:T01} lists the number of galaxies for each sample.
Given the correlation between galaxy morphology and local environment, we also consider 
galaxy samples defined in terms of their morphological types. 
Besides, the merger histories of galaxies with different morphologies could also have 
an impact on possible foregrounds since their gaseous surrounding medium could have very different 
astrophysical properties.
Fig.~\ref{fig:F01} shows the number of galaxies as a function of redshift for early (Sa) and 
late (Sb, Sb~+~Sc~+~Sd) sub-types of Spirals, 
as well as Elliptical galaxies up to $cz = 4500~\rm{km~s^{-1}}$ (panel a).
We notice that the redshift distributions are comparable so 
we can exclude any bias associated with different sample depths. We also present the distributions of 
angular (panel b) and spatial (panel c) sizes for the same morphological types as in panel (a).
Although we cannotice that Spiral galaxies are a little more extended than Elliptical galaxies, there are 
no significant differences.
The vertical line in panel (c) marks the median value of 8.5~Kpc, used as the distinction between large 
and small Spiral galaxies.
In panel (d), we show the distribution of angular distances to the fifth neighbour galaxy within 
$cz \leq 500~\rm{km~s^{-1}}$, and the corresponding threshold median angular distance of 2.68$^\circ$ that 
we use to divide high and low density galaxy environments for the late-type Spiral galaxy subsamples  Sb~+~Sc~+~Sd.
%

%%%%%%%%%%%%%%%%%%%%%%%%%%%%%%%%%%%%%%%%%
\section{Methods}\label{methods}

In this work, we explore the possibility that the surrounding regions of
nearby galaxies affect systematically the derived temperature of CMB photons.
We stress the fact that even if any such effects were present with a very low amplitude, they could still be detectable by 
stacking a large enough number of galaxies.
Accordingly, we have computed the averaged radial temperature profile $\left\langle \Delta~T \right\rangle(\theta)$ as:
\begin{equation}\label{eq:stacking+dT}
\left\langle \Delta~T \right\rangle(\theta) = \frac{1}{N_{\rm{gxs}}} 
\sum_{k = 1}^{N_{\rm{gxs}}}
\left( \frac{1}{N_{k}} \sum_{i \in C_k}
\Delta~T(k, i)\right),
\end{equation}

where the inner sum is performed over the set $C_k$ of all $N_k$ pixels in the ring $(\theta, \theta+\delta\theta)$, where $\theta$ is the angular distance to the centre of each of the $N_{\rm{gxs}}$ galaxies in the sample.

Under the absence of foregrounds or systematic effects, the profiles should be
statistically consistent with zero.
In order to assess the reliability of the signal we have compared
the resulting distribution of radial temperature profiles to those derived for suitable control samples.
For a given sample, each control sample consists of the same number of centres located at random positions within the 
CMB mask, for which we produce stacked CMB profiles with the same radial bins according to equation \eqref{eq:stacking+dT}.
A similar procedure is applied to the analysis of polarisation maps.

\section{Results}

In this section we present the results from the analysis of temperature and polarisation profiles.
We show that the signal strongly depends on the galaxy types, and propose a fiducial model that reasonably 
explains the general behaviour of these profiles considering that only large Spiral, late-type galaxies produce a 
measurable effect, which is transferred to other galaxy populations due to galaxy clustering.
\subsection{Temperature profiles}\label{profiles}
\begin{figure*}
	\includegraphics[width=\textwidth]{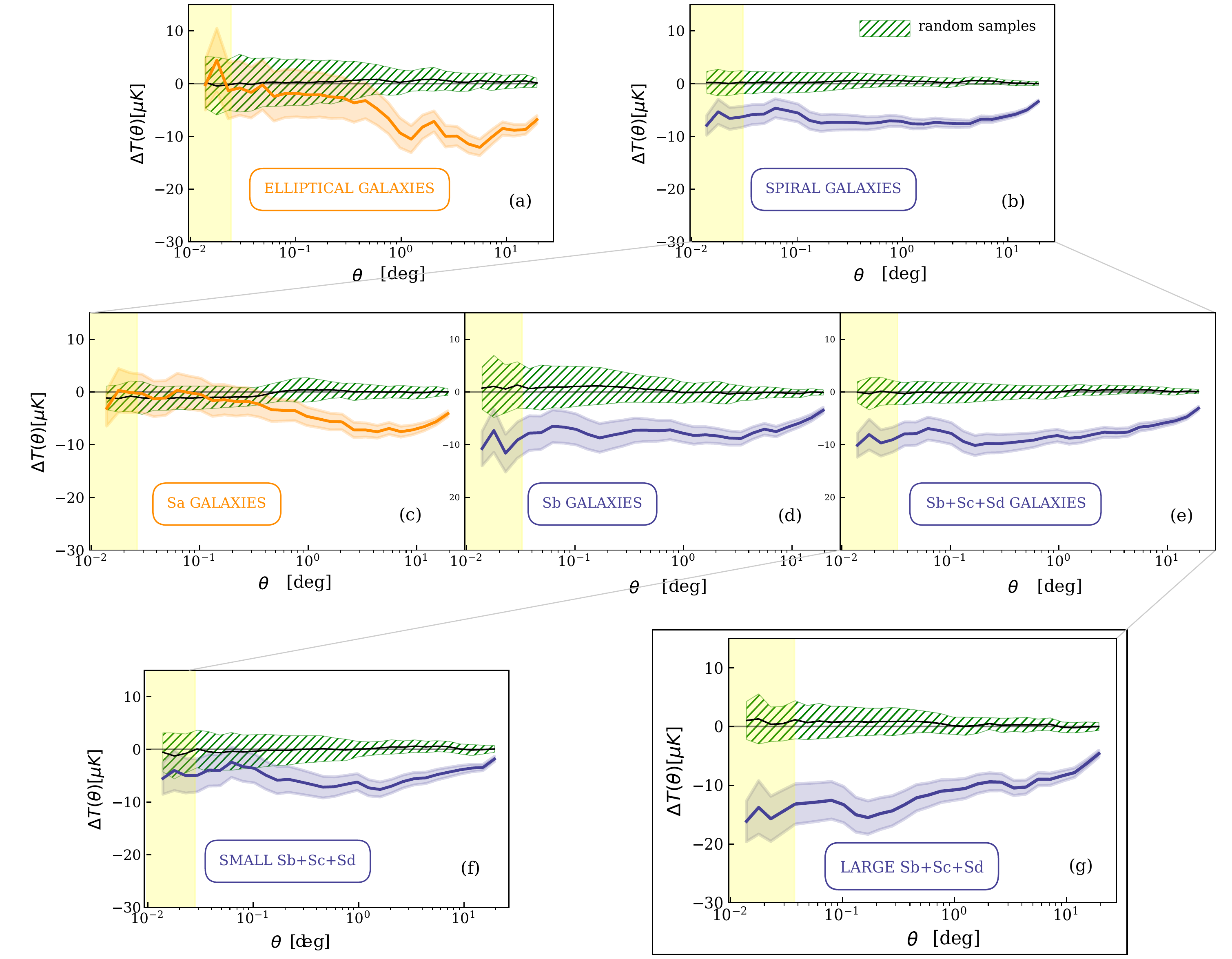}

    \caption{CMB Temperature profiles around galaxy samples defined in terms of their morphological type: Elliptical galaxies (panel a), Spiral galaxies (panel b), Sa (panel c) Sb (panel d) and Sb~+~Sc~+~Sd (panel d). The panels (e) and (f) show the subsamples of small and large Sb~+~Sc~+~Sd galaxies, selected according to the median value of the physical galaxy size shown in panel (d) of Fig.~\ref{fig:F01}. The shaded regions enclosing the profiles correspond to the standard deviations of the mean $\Delta~T(\theta)$ values for the radial bins.
    In all panels, the green shaded regions correspond to the standard deviations of 20 random realisations with the same number of centres than the galaxy data.
    The yellow shaded regions show the angular extent of the galaxies (in terms of $2\times~r_{\rm{ext}}$) up to the mean value for each sample.}
    \label{fig:F02}
\end{figure*}

We have calculated the mean temperature profiles defined in  equation~(\ref{eq:stacking+dT}) for the galaxy samples described in Sec.~\ref{cats}.
In Fig.~\ref{fig:F02}, we present the 
temperature profiles $\Delta~T(\theta)$ around 2MRS galaxies, as a function of the angular separation $\theta$ from the 
galaxy centres for galaxies with $cz \leq 4500~\rm{km~s^{-1}}$.
The shaded regions enclosing the profiles correspond to the standard deviations of the mean $\Delta~T(\theta)$ values for the radial bins.
In all panels, the green shaded regions correspond to the standard deviations of 20 random realisations with the same number of 
centres than the real galaxy data. We have selected 20 realisations given that for a larger number the characterisation of the spread remains unchanged.
The yellow shaded regions show the angular extent of the galaxies (in terms of $2\times~r_{\rm{ext}}$) up to the mean value for each sample.
We have selected this limit to ensure a lack of strong contamination associated with stars and substructure in the inner regions within the optical galaxy extent.
We have analysed the mean radial behaviour of $\Delta~T$ as a function of the
morphological types of the galaxies.
As it can be seen, there is a systematic effect for lower values of $\Delta~T(\theta)$ around nearby galaxies that extends several degrees in the plane of the sky. This systematic effect depends on galaxy morphology, with the latest Hubble type galaxies showing the strongest signal.
Elliptical galaxies (panel a) only present a significant departure from the random samples at large distances from the galaxy centres.
For the total sample of Spiral galaxies (panel b) we can notice a stronger signal. By splitting the sample on Sa (panel c), Sb (panel d) and Sb~+~Sc~+~Sd galaxies (panel e) it is clear that the observed signal is arising from the late-type Spiral galaxies.
To analyse this in further detail, we classified these galaxies according to their size.
We used the median value of the physical size distribution (8.5 Kpc, see panel (c) of Fig.~\ref{fig:F01})
to distinguish between large (panel f) and small (panel g) Sa~+~Sb~+~Sc galaxies. 
This allow us to assess that the origin of the observed $\Delta~T$ decrement is due to the presence of large late-type Spiral galaxies. 
We also see that for all galaxy morphologies there is a slow convergence towards zero at larger separations.

\begin{figure}
	\includegraphics[width=\columnwidth]{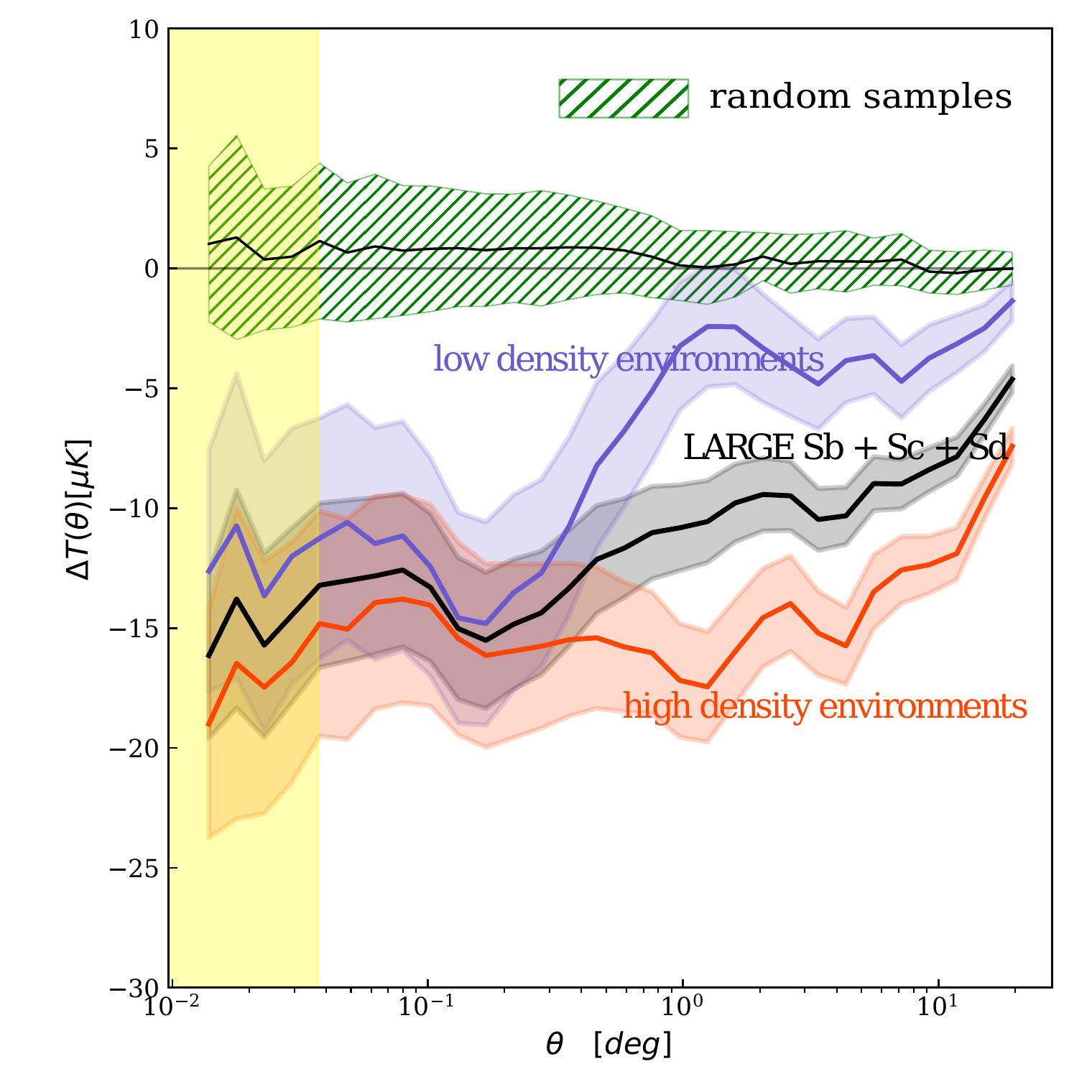}

  \caption{CMB Temperature profiles around Sb~+~Sc~+~Sd galaxies with physical sizes larger than 8.5 Kpc as a function of the environment. The black line shows the total sample. The blue line corresponds to galaxies located in low density environments, and the red line, to galaxies in high density environments. The details of environment classification are described in Section \ref{profiles}.The shaded regions enclosing the profiles correspond to the standard deviations of the mean $\Delta~T(\theta)$ values for the radial bins.
    In all panels, the green shaded regions correspond to the standard deviations of 20 random realisations with the same number of centres than the galaxy data.
    The yellow shaded regions show the angular extent of the galaxies (in terms of $2\times~r_{\rm{ext}}$) up to the mean value for each sample.}
   \label{fig:F03}
\end{figure}

In order to explore the sample of large Sb~+~Sc~+~Sd galaxies into further detail, we have also analysed the dependence of the mean CMB temperature
profile on the local environment of these galaxies. For this aim, we have
computed a suitable proxy of the environmental local density associated to each galaxy given by the angular distance to its fifth neighbouring 
galaxy within $\Delta (cz) = 500~\rm{km~s^{-1}}$. By inspection of panel (d) of Fig.~\ref{fig:F01}, small/large separations correspond to high/low density local environments. The resulting profiles for these galaxy samples are shown in Fig.~\ref{fig:F03}, where it
can be seen that at small angular separations the environmental effect is almost negligible on $\Delta~T$,
while at large separations the environment does play a significant role in the scale of convergence of the signal.

\subsection{Simple model of the effects}\label{models}

The main results of the previous sections consist of statistically significant CMB temperature decrements of $ \sim -15~\mu \rm{K}$ around nearby galaxies with $cz \le 4500~\rm{km~s^{-1}}$. We have also found that this effect is largely dominated by large late-type Spiral galaxies (Sb~+~Sc~+~Sd) since there are no such systematics around elliptical galaxies at close separations, and small late-type galaxies show a marginal effect. These facts suggest a simple model where the presence of systematic large-scale effects of both elliptical and small late-type spiral galaxies could be largely imprinted by those associated only to large late-type Spiral galaxies. The clustering of galaxies could then result in the weaker effect noticed in the other subsamples.
Following this argument, in this section we introduce a simple model that only considers the results of isolated large late-types. As these galaxies are less affected by clustering effects, we can use their temperature profile (see Fig.~\ref{fig:F03}) to infer a suitable model and then apply it to all the large late-type galaxies, which we presume responsible of the observed signal.
We propose a systematic temperature decrement empirically approximated by a law of the form:

\begin{equation}
    T_A(\theta) = 
    \begin{cases}
    -8.66 & \text{if } \theta\le 0.3\text{ deg} \\ 
    7\,\log_{10}(\theta)-5 & \text{if } 0.3\text{ deg}<\theta\le 1\text{ deg} \\ 
    5\,\log_{10}(\theta)-5 & \text{if } 1\text{ deg}<\theta\le 10\text{ deg} \\ 
    0 & \text{if } \theta > 10\text{ deg} \\ 
    \end{cases}
    \label{modelprof}
\end{equation}

We analysed the effects of applying this temperature decrement profile to large late-type Spiral galaxies (Sb~+~Sc~+~Sd with radii larger 
than 8.5 Kpc), regardless of their environment. 
For this aim,  we have analysed a suite of synthetic CMB maps consistent with the \textit{Planck} power spectrum results used as background maps and where we have imposed our model $\Delta~T$ decrement profile around the positions of large late-type Spiral galaxies in the real data. By doing so, we ensure 
that the clustering of the model centre positions is the same as that of the galaxies in the data and therefore it is suitable to test for its effects on other samples. 

\begin{figure*}
	\includegraphics[width=\textwidth]{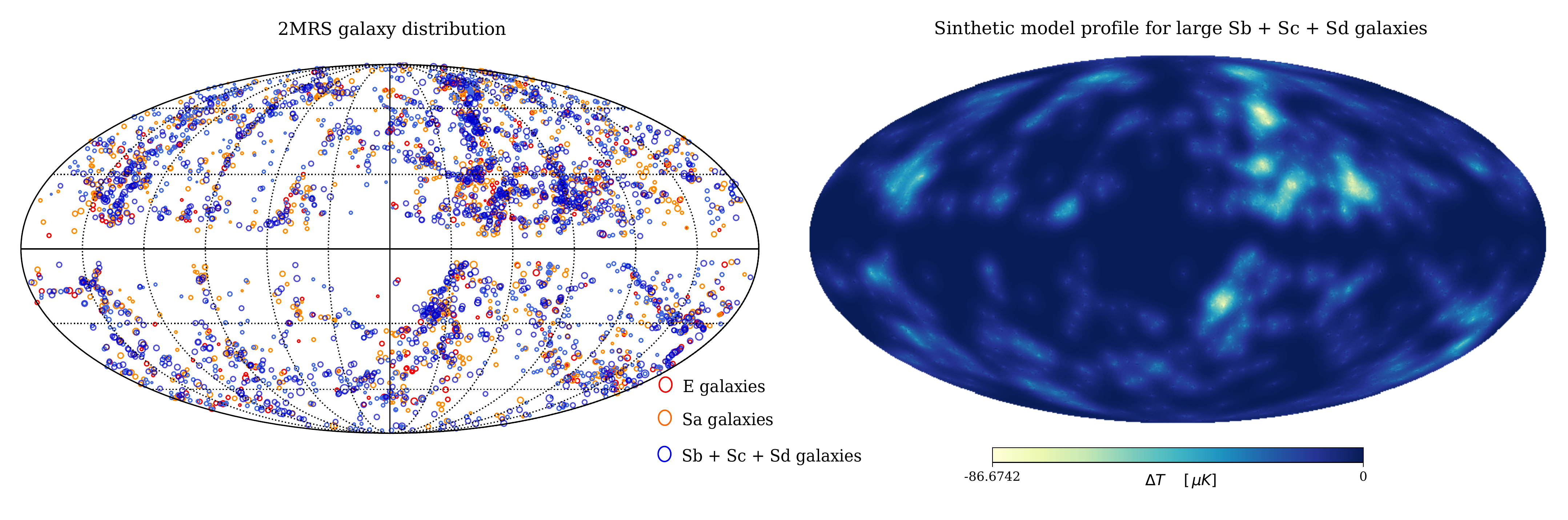}
      \caption{Galaxy distribution and synthetic map. Left panel: Distribution of galaxies from 2MRS in the range $cz <
4500~\rm{km~s^{-1}}$, classified according to their morphological type. Right panel: Synthetic foreground map of $\Delta~T$, obtained by assigning the profile described in Section \ref{models} to the sample of large (radii>8.5~Kpc) Sb~+Sc~+~Sd galaxies. }
    \label{fig:F04}
\end{figure*}

In order to illustrate the extent and strength of the effect, in Fig.~\ref{fig:F04} we show the 2MRS galaxy distribution for our samples, divided by 
their morphology (left), and the resulting map (without the synthetic background) constructed by applying the model profile of the equation (\ref{modelprof}) 
in the positions of the large and isolated Sb~+~Sc~+~Sd galaxies.
Then, by means of equation \ref{eq:stacking+dT} we calculate the mean temperature profiles $\Delta~T(\theta)$.

\begin{figure*}
	\includegraphics[width=\textwidth]{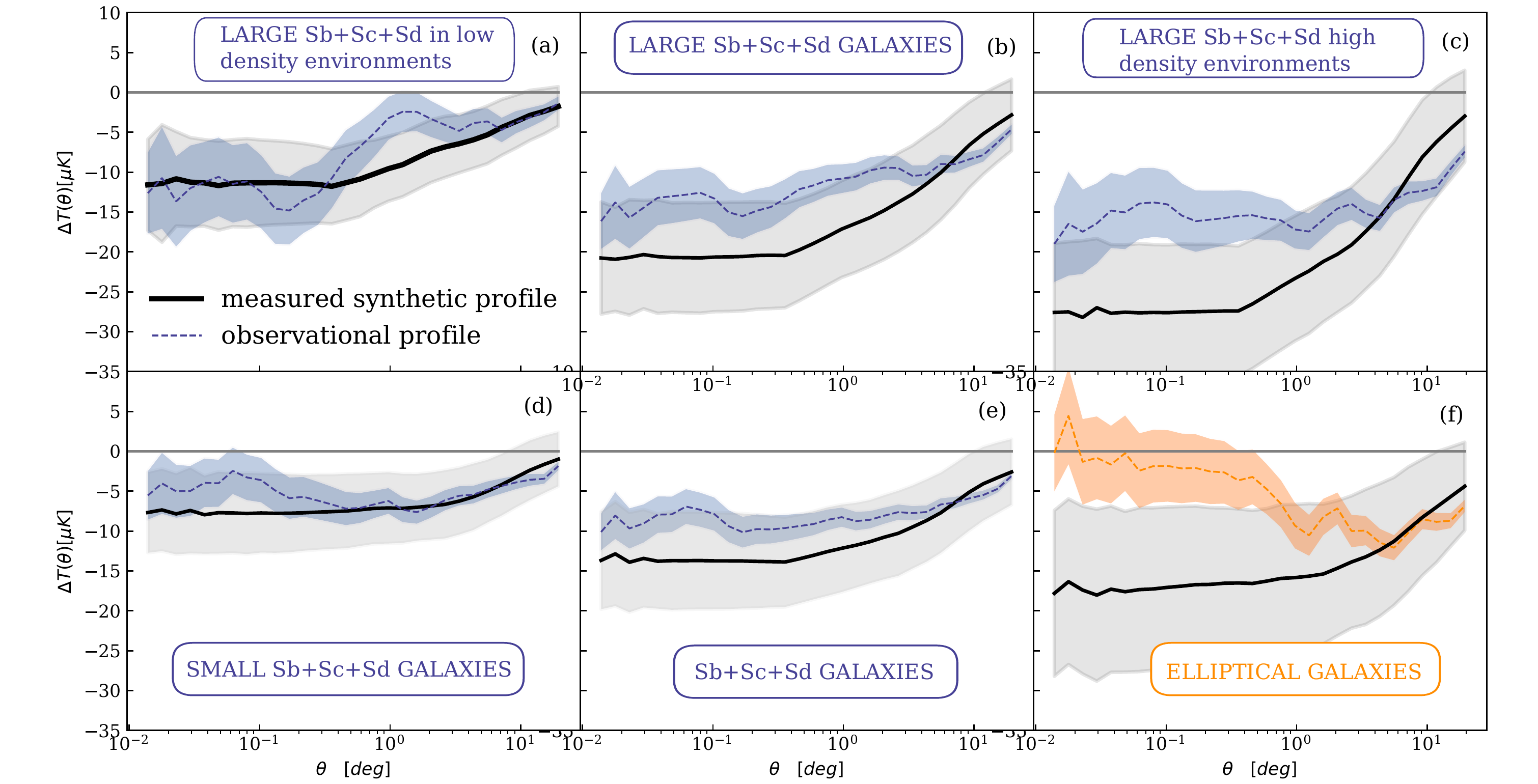}

    \caption{Model temperature profiles obtained for the galaxy samples from the synthetic map described in Section \ref{models}. In each panel, the blue lines correspond to the mean measured profiles from the observational data, and the black lines to the synthetic mean profiles. In both cases, the shaded regions around the profiles correspond to the the standard deviation of the mean $\Delta~T(\theta)$ value for each radial bin.}
   \label{fig:F05}
\end{figure*}

The results for the synthetic profiles are shown in Fig.~\ref{fig:F05}, for a set of 40 realisations with the uncertainty of the mean 
value in each radius bin. We used 40 random realisations for each sample given that for a larger number of realisations the spread remains unchanged. 
For comparison, we overlapped the mean temperature profiles measured in the previous section.  
As we can see, the clustering of galaxies plays an important role as the effect is clearly amplified in high density regions.
As expected, for the case of isolated large late-type Spiral galaxies (panel a), the results are in agreement with the observations. 
However, the success of the simple model relies on the fact that we can reproduce the observed trends for highly-clustered large 
late-type Spirals (panel c), small late-type Spirals (panel d), and Elliptical galaxies (panel f). This fact strongly suggests 
that our ansatz to describe the effects by its association only to large late-type Spiral galaxies is a good approximation to explain 
the origin of the observed signal.

\subsection{Results on polarisation flux} \label{polarisation}

\begin{figure*}
	\includegraphics[width=\textwidth]{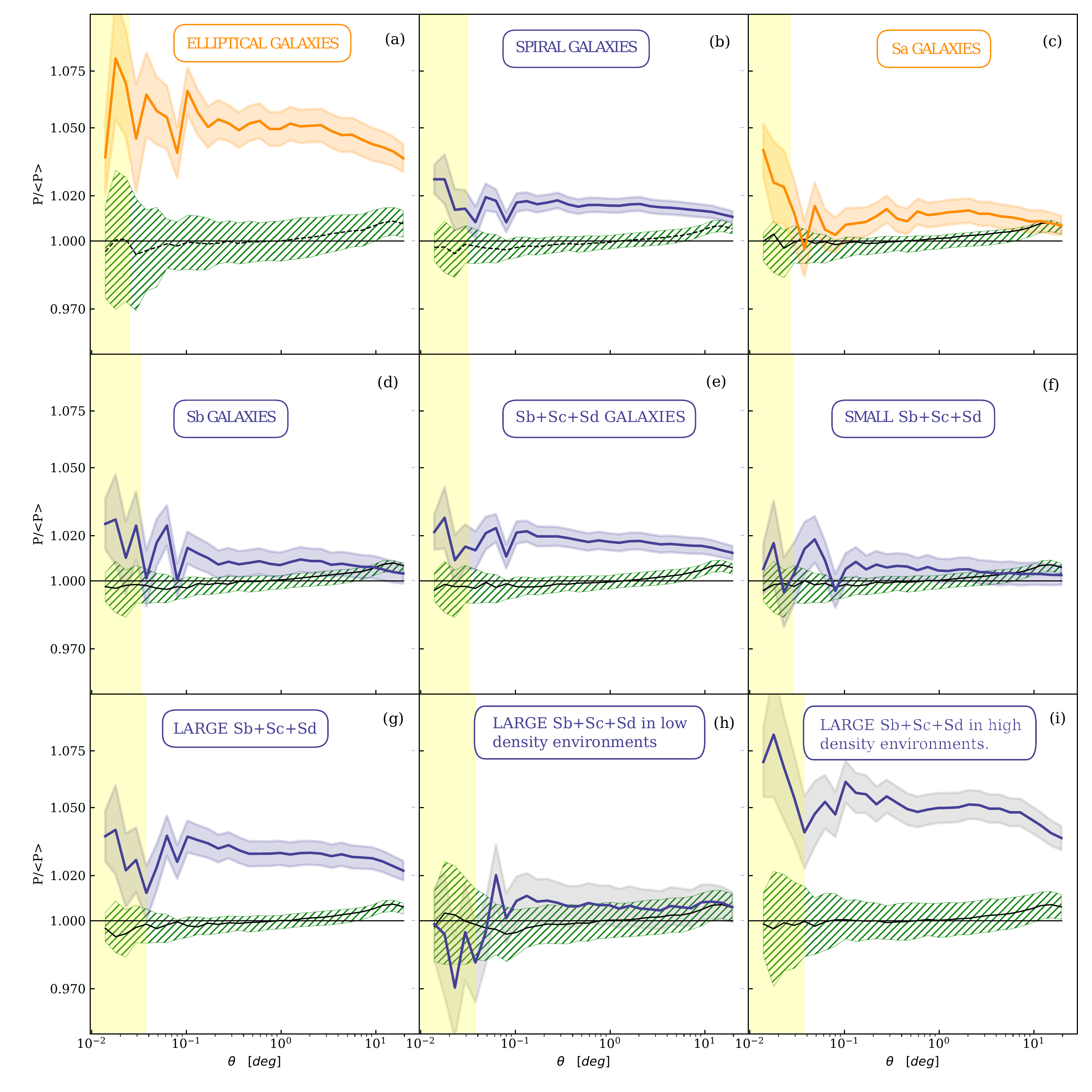}
   \caption{CMB polarisation amplitude (P) profiles around galaxies, according to their morphological type. The normalisation value $<\rm{P}>$ corresponds to the mean polarisation estimated on the total polarisation map. The uncertainties are computed using the same procedures than in the studies on the CMB temperature. The yellow shaded regions show the angular extent of the galaxies (in terms of $2 \times~r_{\rm{ext}}$) up to the mean value for each sample.
   }
   \label{fig:F06}
\end{figure*}

With the aim of further exploring the physical nature associated to
the observed foregrounds, we have analysed possible features in the
polarisation signal.
The polarisation flux amplitude P is defined as $\rm{P}=\sqrt{\rm{Q}^2+\rm{U}^2}$ 
, where Q and U are the Stokes parameters of the polarisation map. By means of the previously defined methods and 
equation (\ref{eq:stacking+dT}), we have calculated the mean polarisation flux profiles around our galaxy samples and their respective uncertainties.
The results are given in Fig.~\ref{fig:F06} where it can be seen a marginal polarised flux for the large late-type Spiral 
galaxies located in high density regions (panel i) and the Elliptical galaxies (panel a).
Unlike the $\Delta~T$ analyses, where the mean value of the temperature maps are of a lower order of 
magnitude with respect to the amplitude of the previously studied signal,  the mean overall P value is  
$<\rm{P}>=35.16~\mu\rm{K}$, almost the same order of 
the measured polarisation signal. Taking this into account, we use this mean value to normalise the estimated 
polarisation fluxes. For this analysis, we used 40 independent random 
realisations for each sample given that for a larger number of realisations the characterisation of the spread remains unchanged.

In spite of the large uncertainties,
this excess of flux has a correspondence with the strongest temperature decrements
obtained in the previous sections. %
Thus, this analysis provides an additional confidence on the reliability
of our results.
The signal for an enhanced polarisation flux surrounding large late-type Spiral galaxies (Sb~+~Sc~+~Sd), gives 
support to the relevant role of dust in the observed temperature foregrounds. 
Furthermore, taking into account the result obtained for Elliptical galaxies, we can also 
infer that the nature of the polarisation signal differs from that detected in temperature. 
This scenario is supported by the presence of polarisation signal and no temperature decrement around Elliptical galaxies.
\section{Summary and discussion}\label{discussion}

In this work we have explored the presence of CMB foregrounds associated with nearby galaxies. %		
For this aim, we have analysed the mean CMB temperature profiles around the positions of 
several samples of low-redshift galaxies, taken from the 2MRS catalogue.
We found that the positions of nearby Spiral galaxies ($cz < 4500~\rm{km~s^{-1}}$) are associated 
with statistically significant CMB temperature decrements that extend to several galaxy radii.
This effect is found using \textit{Planck} CMB temperature anisotropy maps and is also well reproduced in the final 
release of the \textit{WMAP} sky maps (see Appendix~\ref{app_wmap}).

These data-sets, have been subject to data reduction pipelines that implement independent methods to remove 
the foreground contamination. 
Nevertheless, in order to  address possible systematic effects in foreground remotion procedures in the data, we have also analysed directly the flux  mean profiles in different frequencies. The results for the samples explored along the paper have remarkably the same patterns as those derived from the SMICA maps. This further assures that our findings cannot be associated with residuals originated in foreground extraction applied to the data.
Although there are several known sources of foregrounds that are removed at high confidence
level through these pipelines, our findings suggest that some sources of contamination from the nearby universe are still present in the CMB data and deserve attention in order achieve better understanding of the possible presence of more distant foreground effects. 

For each of the ten bins between 1$^{\circ}$ and 10$^{\circ}$. the significance of the effect for the sample of large late-type 
Spiral galaxies (Sb+Sc+Sd) is at least at the 4$\sigma$ level. Thus, our results are reliable at a high confidence level.
Generally, at separations greater than $1^{\circ}$ from galaxies, the samples show a similar $\Delta~T$ signal (approximately $\Delta~T \simeq -10~\mu \rm{K}$)
for both Elliptical and Spiral galaxies. However, closer to the galaxies, only Spiral galaxies show this systematic mean decrements.
A further analysis considering only sub-samples with late morphological sub-types (Sb~+~Sc~+~Sd) shows that these galaxies are associated with the strongest 
decrements, accounting for approximately $\Delta~T \simeq -15~\mu \rm{K}$.
A more detailed exploration considering the properties of the galaxy sample also reveals that the largest galaxies
show the strongest signals, as well as those galaxies with more close neighbours for which we find a difference in the $ \Delta~T$ signal at larger separations 
(this behaviour is shown in Fig.~\ref{fig:F03}). 

Overall, the presence of foregrounds associated with the galaxy samples analysed here, together with the features extracted from the measured signal,
suggest that astrophysical processes take place on large intergalactic scales.
In particular, the fact that the strongest effects are obtained
for large late-type Spirals with closer neighbours could be a hint for its relation to tidal interactions of neighbour galaxies with the large central Spiral galaxy. As a result of the interaction
between galaxies and the intergalactic medium coupled to ram pressure stripping effects, a significant amount of
gas from approaching galaxies may get stripped in their orbital
turnover. To this regard it is also worthwhile to mention that simulations suggest that metallic gas removed from the
central regions of galaxies may gain sufficient energy to get extended
beyond the group virial radius, reaching several hundreds of kpc (Rodriguez et al., accepted for publication in MNRAS).  
In this scenario, orbiting satellites affected by tidal forces and ram
pressure stripping may suitably spread dust at large distances from
galaxies. 
Large Elliptical and Spiral galaxies should have
a similar CMB effect at large scales, as observed.
%
%Closer to the galaxies, we would expect a lack of effects for Elliptical galaxies,
%where the interaction-induced  mechanisms proposed here should have occurred much earlier leaving enough time to spread away the removed material \AGS{(Rodriguez et al., in prep.)}. 
We acknowledge that the precise nature of the interaction between CMB photons and the intergalactic medium surrounding the galaxies cannot be fully derived from our analysis.
The ISW effect is a common source of discussion within the so-called anomalies of the $\Lambda$CDM on large scales (see for example \citet{Kovacs:2020lqm}).
In \citet{Nadathur:2011iu}, a positive mean temperature was reported around overdensity regions using red luminous galaxies at redshifts greater than $z = 0.4$. On the contrary, our findings correspond to a mean temperature decrement 
around these particular overdense regions.
However, taking into account the order of magnitude of the signal, and the fact that $\Delta~T$ is negative, we can rule out well known foreground sources often discussed in the literature, where an approach combining CMB measurements and galaxy samples are used.
For instance, due to the large difference between Spiral and Elliptical galaxies, our detection cannot be directly ascribed to the ISW effect even though the order of magnitude of the expected signal is within the values presented here.
Also, as shown in \citet{Makiya2018}, the thermal Sunyaev-Zeldovich effect is present around locally bright galaxies. 
To explore the effects of this phenomenon, we have considered
in Appendix \ref{app_sz} the $\Delta T(\theta)$ profiles for our samples, 
computed over the SMICA map where Sunyaev-Zeldovich sources have been projected out.
This allow us to confirm that the present results cannot be attributed to the thermal SZ effect due to known sources.
%XXX
%
We also stress the fact that for all samples with significant temperature
decrements (those of Fig.~\ref{fig:F03}), the temperature profiles exhibit a feature at 5 to 10 arcmin, which corresponds to approximately 100~kpc.
This scale seems to divide the inner regime, dominated by galaxies in
isolation, and the large-scale behaviour, strongly dependent on
environment, as is most clearly seen in Fig.~\ref{fig:F03}.

In a forthcoming paper (Boero et al., in preparation) we will address 
the possible effects on the CMB power spectrum and
its implications for cosmological parameter inference. These studies
may be relevant for the accuracy of the derived cosmological parameters from the CMB.

\section*{Data availability}
The datasets used in this article were derived from sources in the public domain. 
Data products and observations from $Planck$, \url{https://pla.esac.esa.int}.
Data products from the Two Micron All Sky Survey,
\url{http://tdc-www.cfa.harvard.edu/2mrs}.
Data products from WMAP mission \url{https://lambda.gsfc.nasa.gov/product/wmap/dr5/m_products.html}.

\section*{Acknowledgements}
%

%
%XXX
We thank Prof. James Peebles and Prof. Eiichiro Komatsu for their useful comments on this work.
%XXX
%

This work was partially supported by the Consejo Nacional de
Investigaciones Cient\'{\i}ficas y T\'ecnicas (CONICET), and the
Secretar\'{\i}a de Ciencia y Tecnolog\'{\i}a (SeCyT), Universidad Nacional de
C\'ordoba, Argentina.
This work is based on observations obtained with Planck (http://www.esa.int/Planck), an ESA science mission with instruments and contributions directly funded by ESA Member States, NASA, and Canada.
Some of the results in this paper have been derived using the HEALPix (K.M. Górski et al., 2005, ApJ, 622, p759) package.
Plots were made using Python software and post-processed with
Inkscape.
%
%Most of the data analyisis was carried out using SciPy for numerical sampling of the statistical distributions involved in data analysis,
%and Numpy for numerical linear algebra. 
%
% This research has made use of NASA's Astrophysics Data System. 
%
% All the data underlying this article are publicly available.

%%%%%%%%%%%%%%%%%%%%%  BIBLIOGRAPHY  %%%%%%%%%%%%%

\bibliographystyle{mnras}
\bibliography{bibliography}

\begin{thebibliography}{}
\makeatletter
\relax
\def\mn@urlcharsother{\let\do\@makeother \do\$\do\&\do\#\do\^\do\_\do\%\do\~}
\def\mn@doi{\begingroup\mn@urlcharsother \@ifnextchar [ {\mn@doi@}
  {\mn@doi@[]}}
\def\mn@doi@[#1]#2{\def\@tempa{#1}\ifx\@tempa\@empty \href
  {http://dx.doi.org/#2} {doi:#2}\else \href {http://dx.doi.org/#2} {#1}\fi
  \endgroup}
\def\mn@eprint#1#2{\mn@eprint@#1:#2::\@nil}
\def\mn@eprint@arXiv#1{\href {http://arxiv.org/abs/#1} {{\tt arXiv:#1}}}
\def\mn@eprint@dblp#1{\href {http://dblp.uni-trier.de/rec/bibtex/#1.xml}
  {dblp:#1}}
\def\mn@eprint@#1:#2:#3:#4\@nil{\def\@tempa {#1}\def\@tempb {#2}\def\@tempc
  {#3}\ifx \@tempc \@empty \let \@tempc \@tempb \let \@tempb \@tempa \fi \ifx
  \@tempb \@empty \def\@tempb {arXiv}\fi \@ifundefined
  {mn@eprint@\@tempb}{\@tempb:\@tempc}{\expandafter \expandafter \csname
  mn@eprint@\@tempb\endcsname \expandafter{\@tempc}}}

\bibitem[\protect\citeauthoryear{Afshordi, Loh  \& Strauss}{Afshordi
  et~al.}{2004}]{Afshordi2004}
Afshordi N.,  Loh Y.-S.,   Strauss M.~A.,  2004, \mn@doi [Phys. Rev. D]
  {10.1103/PhysRevD.69.083524}, 69, 083524

\bibitem[\protect\citeauthoryear{Aghanim, Akrami, Alves, Ashdown  \& et
  al.}{Aghanim et~al.}{2020a}]{Aghanim2020b}
Aghanim N.,  Akrami Y.,  Alves M.~I.,  Ashdown M.,   et al. J.~A.,  2020a,
  \mn@doi [Astronomy and Astrophysics] {10.1051/0004-6361/201833885}, 641, 24

\bibitem[\protect\citeauthoryear{Aghanim, Akrami, Arroja, Ashdown  \& et
  al.}{Aghanim et~al.}{2020b}]{Aghanim2020c}
Aghanim N.,  Akrami Y.,  Arroja F.,  Ashdown M.,   et al. J.~A.,  2020b,
  \mn@doi [Astronomy and Astrophysics] {10.1051/0004-6361/201833880}, 641, A1

\bibitem[\protect\citeauthoryear{Aghanim, Akrami, Ashdown, Aumont  \& et
  al.}{Aghanim et~al.}{2020c}]{Aghanim2020a}
Aghanim N.,  Akrami Y.,  Ashdown M.,  Aumont J.,   et al. C.~B.,  2020c,
  \mn@doi [Astronomy and Astrophysics] {10.1051/0004-6361/201832909}, 641, A3

\bibitem[\protect\citeauthoryear{{Alpher}, {Bethe}  \& {Gamow}}{{Alpher}
  et~al.}{1948}]{Alpher1948}
{Alpher} R.~A.,  {Bethe} H.,   {Gamow} G.,  1948, \mn@doi [Physical Review]
  {10.1103/PhysRev.73.803}, \href
  {https://ui.adsabs.harvard.edu/abs/1948PhRv...73..803A} {73, 803}

\bibitem[\protect\citeauthoryear{{Bennett} et~al.,}{{Bennett}
  et~al.}{2003}]{Bennet:2003b}
{Bennett} C.~L.,  et~al., 2003, \mn@doi [\apjs] {10.1086/377252}, \href
  {https://ui.adsabs.harvard.edu/abs/2003ApJS..148...97B} {148, 97}

\bibitem[\protect\citeauthoryear{{Bennett} et~al.,}{{Bennett}
  et~al.}{2013a}]{2013ApJS..208...20B}
{Bennett} C.~L.,  et~al., 2013a, \mn@doi [\apjs] {10.1088/0067-0049/208/2/20},
  \href {https://ui.adsabs.harvard.edu/abs/2013ApJS..208...20B} {208, 20}

\bibitem[\protect\citeauthoryear{{Bennett} et~al.,}{{Bennett}
  et~al.}{2013b}]{Bennett2013}
{Bennett} C.~L.,  et~al., 2013b, \mn@doi [\apjs] {10.1088/0067-0049/208/2/20},
  \href {https://ui.adsabs.harvard.edu/abs/2013ApJS..208...20B} {208, 20}

\bibitem[\protect\citeauthoryear{{Blanton} et~al.,}{{Blanton}
  et~al.}{2005}]{NYUVAGC}
{Blanton} M.~R.,  et~al., 2005, \mn@doi [\aj] {10.1086/429803}, \href
  {https://ui.adsabs.harvard.edu/abs/2005AJ....129.2562B} {129, 2562}

\bibitem[\protect\citeauthoryear{{Cao}, {Chu}  \& {Fang}}{{Cao}
  et~al.}{2006}]{Cao2006}
{Cao} L.,  {Chu} Y.-Q.,   {Fang} L.-Z.,  2006, \mn@doi [\mnras]
  {10.1111/j.1365-2966.2006.10277.x}, \href
  {https://ui.adsabs.harvard.edu/abs/2006MNRAS.369..645C} {369, 645}

\bibitem[\protect\citeauthoryear{{De Paolis} et~al.,}{{De Paolis}
  et~al.}{2019}]{DePaolis:2019hnq}
{De Paolis} F.,  et~al., 2019, \mn@doi [\aap] {10.1051/0004-6361/201936327},
  \href {https://ui.adsabs.harvard.edu/abs/2019A&A...629A..87D} {629, A87}

\bibitem[\protect\citeauthoryear{{Dicke}, {Peebles}, {Roll}  \&
  {Wilkinson}}{{Dicke} et~al.}{1965}]{Dicke1965}
{Dicke} R.~H.,  {Peebles} P.~J.~E.,  {Roll} P.~G.,   {Wilkinson} D.~T.,  1965,
  \mn@doi [\apj] {10.1086/148306}, \href
  {https://ui.adsabs.harvard.edu/abs/1965ApJ...142..414D} {142, 414}

\bibitem[\protect\citeauthoryear{Dodelson}{Dodelson}{2003}]{Dodelson2003}
Dodelson S.,  2003, {Modern Cosmology}.
Academic Press, Amsterdam

\bibitem[\protect\citeauthoryear{Francis \& Peacock}{Francis \&
  Peacock}{2009a}]{Francis2009a}
Francis C.~L.,  Peacock J.~A.,  2009a, \mn@doi [\mnras]
  {10.1111/j.1365-2966.2010.16278.x}, 406, 2

\bibitem[\protect\citeauthoryear{Francis \& Peacock}{Francis \&
  Peacock}{2009b}]{Francis2009b}
Francis C.~L.,  Peacock J.~A.,  2009b, \mn@doi [\mnras]
  {10.1111/j.1365-2966.2010.16866.x}, 406, 14

\bibitem[\protect\citeauthoryear{Gamow}{Gamow}{1946}]{Gamow1946}
Gamow G.,  1946, \mn@doi [Physical Review] {10.1103/PhysRev7.0.572}, 70, 572

\bibitem[\protect\citeauthoryear{{Hern{\'a}ndez-Monteagudo}, {Trac}, {Verde}
  \& {Jimenez}}{{Hern{\'a}ndez-Monteagudo}
  et~al.}{2006}]{hernandez_thermal_2006}
{Hern{\'a}ndez-Monteagudo} C.,  {Trac} H.,  {Verde} L.,   {Jimenez} R.,  2006,
  \mn@doi [\apjl] {10.1086/510123}, \href
  {https://ui.adsabs.harvard.edu/abs/2006ApJ...652L...1H} {652, L1}

\bibitem[\protect\citeauthoryear{{Huchra} et~al.,}{{Huchra}
  et~al.}{2012}]{Huchra2012_2MRS}
{Huchra} J.~P.,  et~al., 2012, \mn@doi [\apjs] {10.1088/0067-0049/199/2/26},
  \href {https://ui.adsabs.harvard.edu/abs/2012ApJS..199...26H} {199, 26}

\bibitem[\protect\citeauthoryear{{Jarrett}, {Chester}, {Cutri}, {Schneider},
  {Skrutskie}  \& {Huchra}}{{Jarrett} et~al.}{2000}]{Jarret:2000}
{Jarrett} T.~H.,  {Chester} T.,  {Cutri} R.,  {Schneider} S.,  {Skrutskie} M.,
   {Huchra} J.~P.,  2000, \mn@doi [\aj] {10.1086/301330}, \href
  {https://ui.adsabs.harvard.edu/abs/2000AJ....119.2498J} {119, 2498}

\bibitem[\protect\citeauthoryear{{Kov{\'a}cs}, {Beck}, {Szapudi}, {Csabai},
  {R{\'a}cz}  \& {Dobos}}{{Kov{\'a}cs} et~al.}{2020}]{Kovacs:2020lqm}
{Kov{\'a}cs} A.,  {Beck} R.,  {Szapudi} I.,  {Csabai} I.,  {R{\'a}cz} G.,
  {Dobos} L.,  2020, \mn@doi [\mnras] {10.1093/mnras/staa2631}, \href
  {https://ui.adsabs.harvard.edu/abs/2020MNRAS.499..320K} {499, 320}

\bibitem[\protect\citeauthoryear{{Makiya}, {Ando}  \& {Komatsu}}{{Makiya}
  et~al.}{2018}]{Makiya2018}
{Makiya} R.,  {Ando} S.,   {Komatsu} E.,  2018, \mn@doi [\mnras]
  {10.1093/mnras/sty2031}, \href
  {https://ui.adsabs.harvard.edu/abs/2018MNRAS.480.3928M} {480, 3928}

\bibitem[\protect\citeauthoryear{Nadathur, Hotchkiss  \& Sarkar}{Nadathur
  et~al.}{2012}]{Nadathur:2011iu}
Nadathur S.,  Hotchkiss S.,   Sarkar S.,  2012, \mn@doi [JCAP]
  {10.1088/1475-7516/2012/06/042}, 06, 042

\bibitem[\protect\citeauthoryear{{Peebles}}{{Peebles}}{1980}]{Peebles1980}
{Peebles} P.~J.~E.,  1980, {The large-scale structure of the universe}.
Princeton University Press, 1980.~435 p.

\bibitem[\protect\citeauthoryear{{Penzias} \& {Wilson}}{{Penzias} \&
  {Wilson}}{1965}]{Penzias1965}
{Penzias} A.~A.,  {Wilson} R.~W.,  1965, \mn@doi [\apj] {10.1086/148307}, \href
  {https://ui.adsabs.harvard.edu/abs/1965ApJ...142..419P} {142, 419}

\bibitem[\protect\citeauthoryear{{Planck Collaboration} et~al.,}{{Planck
  Collaboration} et~al.}{2013}]{Planck_XI}
{Planck Collaboration} et~al., 2013, \mn@doi [\aap]
  {10.1051/0004-6361/201220941}, \href
  {https://ui.adsabs.harvard.edu/abs/2013A&A...557A..52P} {557, A52}

\bibitem[\protect\citeauthoryear{{Planck Collaboration} et~al.,}{{Planck
  Collaboration} et~al.}{2014}]{PlanckXII_2013}
{Planck Collaboration} et~al., 2014, \mn@doi [\aap]
  {10.1051/0004-6361/201321580}, \href
  {https://ui.adsabs.harvard.edu/abs/2014A&A...571A..12P} {571, A12}

\bibitem[\protect\citeauthoryear{{Planck Collaboration} et~al.,}{{Planck
  Collaboration} et~al.}{2016}]{PlanckIX_2015}
{Planck Collaboration} et~al., 2016, \mn@doi [\aap]
  {10.1051/0004-6361/201525936}, \href
  {https://ui.adsabs.harvard.edu/abs/2016A&A...594A...9P} {594, A9}

\bibitem[\protect\citeauthoryear{{Planck Collaboration} et~al.,}{{Planck
  Collaboration} et~al.}{2020}]{planck2018IV}
{Planck Collaboration} et~al., 2020, \mn@doi [\aap]
  {10.1051/0004-6361/201833881}, \href
  {https://ui.adsabs.harvard.edu/abs/2020A&A...641A...4P} {641, A4}

\bibitem[\protect\citeauthoryear{{Rassat}, {Land}, {Lahav}  \&
  {Abdalla}}{{Rassat} et~al.}{2007}]{Rassat2006}
{Rassat} A.,  {Land} K.,  {Lahav} O.,   {Abdalla} F.~B.,  2007, \mn@doi
  [\mnras] {10.1111/j.1365-2966.2007.11538.x}, \href
  {https://ui.adsabs.harvard.edu/abs/2007MNRAS.377.1085R} {377, 1085}

\bibitem[\protect\citeauthoryear{{Rassat}, {Starck}  \& {Dup{\'e}}}{{Rassat}
  et~al.}{2013}]{Rassat2013}
{Rassat} A.,  {Starck} J.~L.,   {Dup{\'e}} F.~X.,  2013, \mn@doi [\aap]
  {10.1051/0004-6361/201219793}, \href
  {https://ui.adsabs.harvard.edu/abs/2013A&A...557A..32R} {557, A32}

\bibitem[\protect\citeauthoryear{{de Vaucouleurs}}{{de
  Vaucouleurs}}{1963}]{devaucouleurs_revised_1963}
{de Vaucouleurs} G.,  1963, \mn@doi [\apjs] {10.1086/190084}, \href
  {https://ui.adsabs.harvard.edu/abs/1963ApJS....8...31D} {8, 31}

\makeatother
\end{thebibliography}

%%%%%%%%%%%%%%%%% APPENDICES %%%%%%%%%%%%%%%%%%%%%
\appendix
%\section*{Appendices}

\section{Redshift dependence}\label{app_z}
\begin{figure}
	\includegraphics[width=\columnwidth]{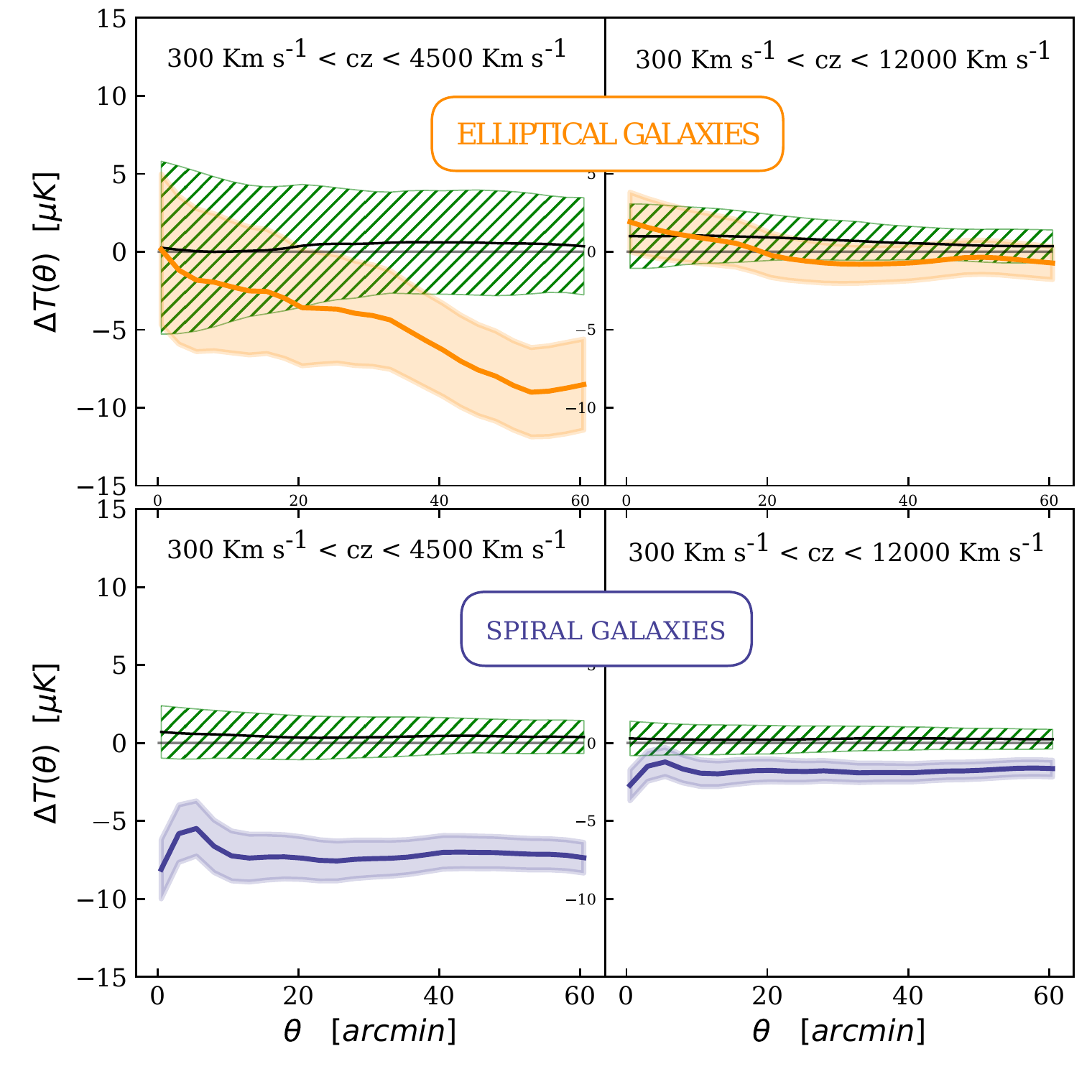}

    \caption{Temperature profiles around 2MRS galaxies, as a function of the maximum redshift of the samples. The upper panels correspond to Elliptical galaxies and the lower panels to Spiral galaxies.
    Right and left panels correspond to the samples up to $300~\rm{km~s^{-1}}<cz<4500~\rm{km~s^{-1}}$ and $300~\rm{km~s^{-1}}<cz<12000~\rm{km~s^{-1}}$ respectively. 
    The shaded regions enclosing the profiles correspond to the standard deviations of the mean $\Delta~T(\theta)$ values for the radial bins.
    In all panels, the green shaded regions correspond to the standard deviations of 20 random realisations with the same number of centres than the galaxy data.
    }%
    \label{fig:FA1}
\end{figure}

We have studied the impact of the depth of the galaxy samples from the 2MRS data.
We have used the methods described in Section \ref{methods} to obtain the $\Delta~T(\theta)$ 
profile in the \textit{Plank} data, for the samples of Elliptical and Spiral galaxies with 
$300~\rm{km~s^{-1}}<cz<12000~\rm{km~s^{-1}}$ from Table \ref{tab:T01}, and compared these results with the ones  
described  in Section \ref{profiles}.
In Fig.~\ref{fig:FA1} we show the temperature profiles as a function of angular distance to the galaxy
centre for the two samples.
The upper panels correspond to Elliptical galaxies and the lower
panels to Spiral galaxies.
Left and right panels correspond to the samples in the ranges
$300 ~{\rm km~s^{-1}}<cz<4500~{\rm km~s^{-1}}$ and $300~{\rm km~s^{-1}}<cz<12000~{\rm km~s^{-1}}$, respectively.
The shaded regions enclosing the profiles correspond to the standard deviations of the mean $\Delta~T(\theta)$ values for the radial bins.
In all panels, the green shaded regions correspond to the standard deviations of 20 random realisations with the same number of centres than the galaxy data.
For simplicity, we have restricted our analysis to lower distances from the galaxy centres, where we have found a stronger signal.
We can see that the selection of a deeper sample naturally has impact on the accuracy of the determination of the effect (as inferred from the comparison of the shaded errors).
Nevertheless, the signal almost vanishes when we use more distant galaxies. This suggests that the foreground effect is due to nearby large Spiral galaxies.

\section{Analysis of \textit{WMAP} data} \label{app_wmap}
\begin{figure}
	\includegraphics[width=\columnwidth]{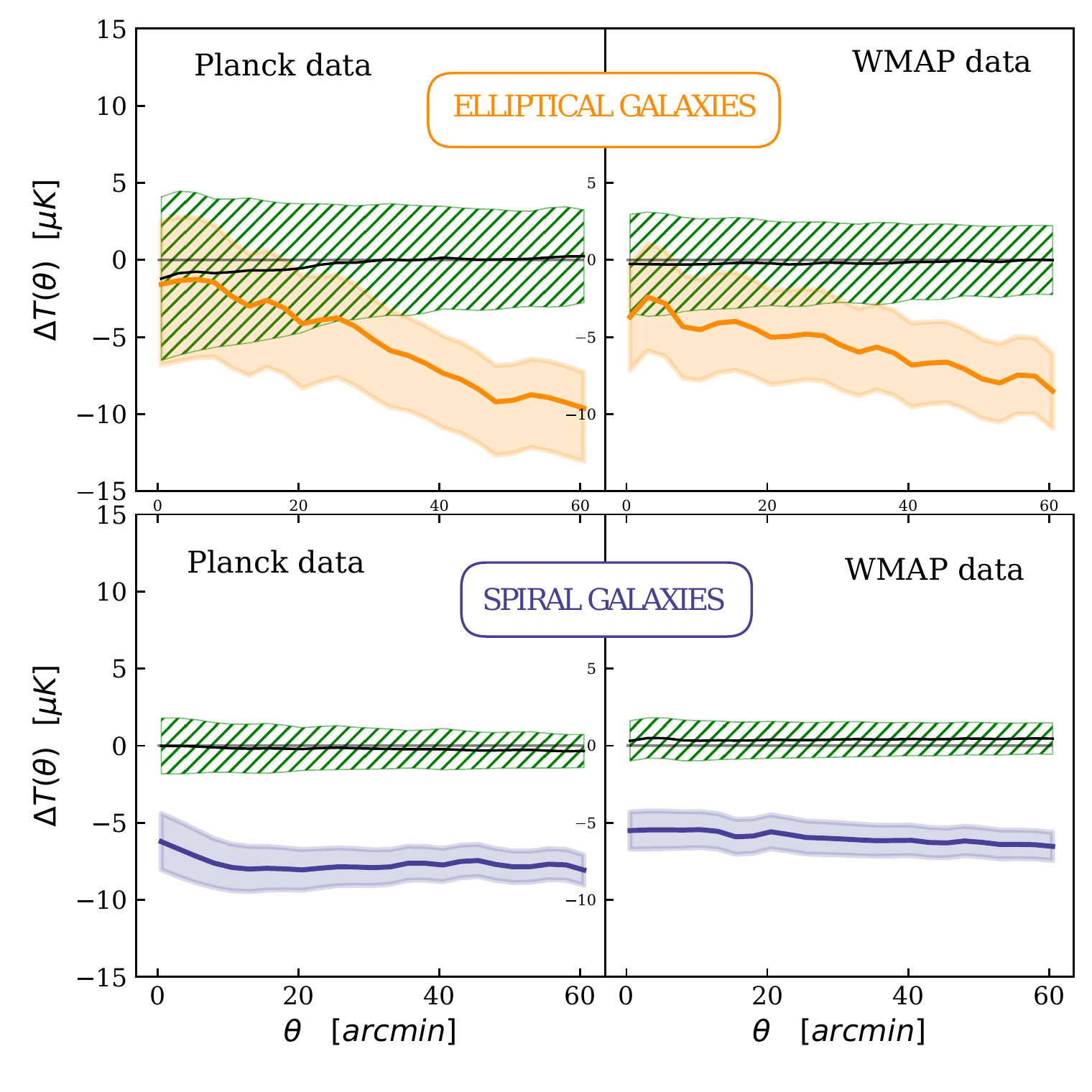}
   \caption{Temperature profiles around 2MRS galaxies in the range $cz\le 4500~\rm{km~s^{-1}}$ for \textit{Plank} and \textit{WMAP} data. The right panels correspond to the $\Delta~T(\theta)$ profile for \textit{Planck} data, and the left panels to the profiles for \textit{WMAP} data. The upper panels show the results for the Elliptical galaxies and the lower panels for the Spiral galaxy samples. 
   The shaded regions enclosing the profiles correspond to the standard deviations of the mean $\Delta~T(\theta)$ values for the radial bins.
    In all panels, the green shaded regions correspond to the standard deviations of 20 random realisations with the same number of centres than the galaxy data.
    Also, both CMB maps are normalized to its corresponding zero point and were used in a resulution of $\rm{N}_{\rm{side}}=512$ for a fair comparison.}
   \label{fig:FA2}
\end{figure}

Here, we present the results of a similar cross-correlation analysis to the one described in Sec.~\ref{profiles} 
applied to \textit{WMAP} data \citep{2013ApJS..208...20B}.
This comparison allows us to assess a possible dependence of our results on the different data reduction pipelines 
and is a strong statistically independent test for the reported effects.
In this appendix we show the results of mean temperature
profiles around the samples of Elliptical and Spiral galaxies. 
In Fig.~\ref{fig:FA2} we show the signal of the mean temperature profiles around the samples of Elliptical and Spiral galaxies, up to $cz\le 4500~\rm{km~s^{-1}}$ in both Planck and WMAP cleaned maps.
It should be noticed that although the pixel size resolution of \textit{WMAP} temperature map is lower than the resolution of \textit{Planck}, it is still suitable for our analysis.
In order to make a fair comparison, both CMB maps have been normalized to their corresponding zero point,
and were used in a common $\rm{N}_{\rm{side}}=512$ resolution, which results in a pixel area  $\Omega_{\rm{p}} = 1.699
\times 10^5\rm{arcsec}^2$. 
As it can be seen, we recover similar signals in \textit{WMAP} data giving further support to our
results.

\section{Thermal Sunyaev-Zeldovich effect}\label{app_sz}
\begin{figure*}
	\includegraphics[width=\textwidth]{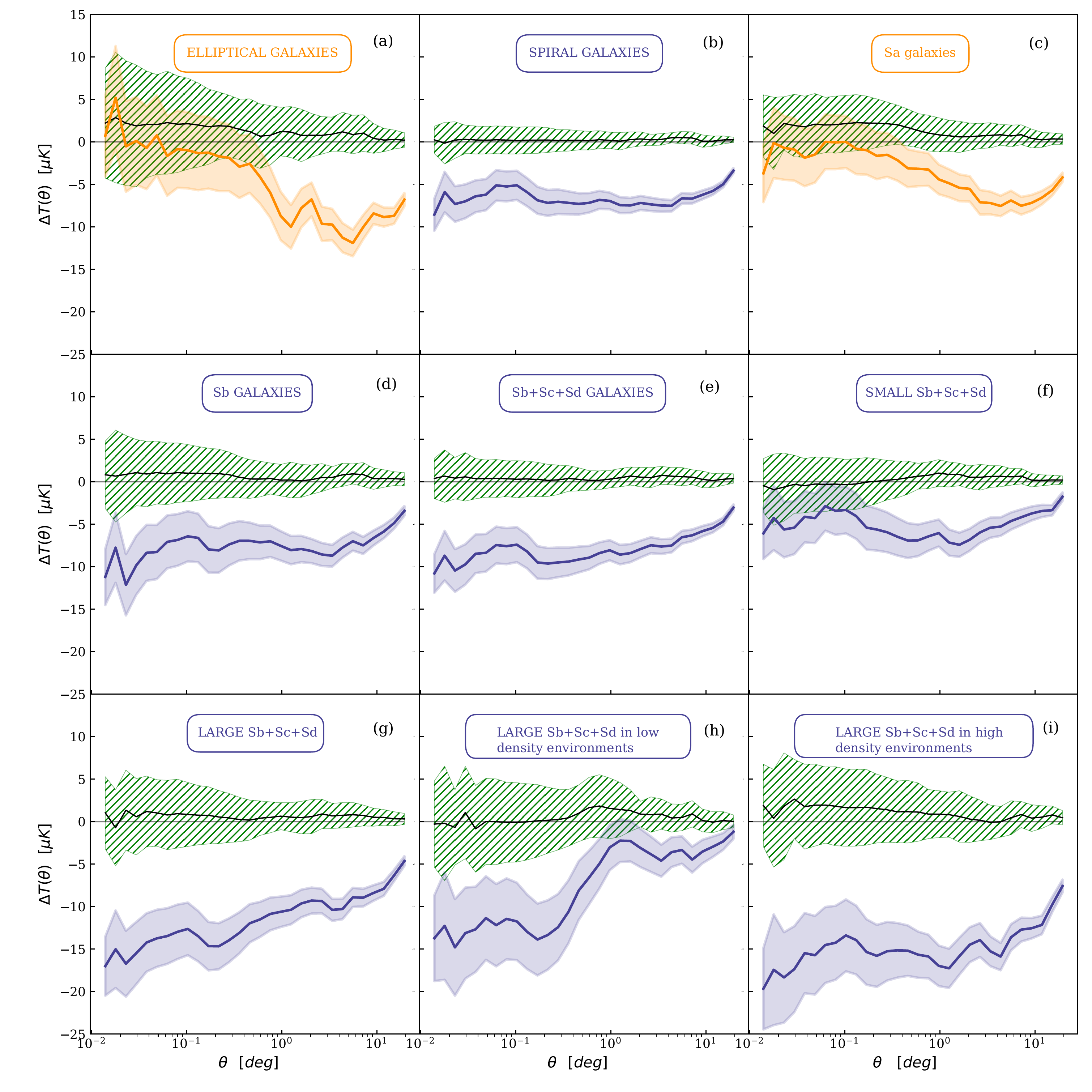}
   \caption{Temperature profiles from the SMICA map where Sunyaev-Zeldovich sources have been projected out, around 2MRS galaxies in the range $cz\le 4500~\rm{km~s^{-1}}$ according to their morphological type: Elliptical galaxies (panel a), Spiral galaxies (panel b), Sa (panel c) Sb (panel d) and Sb~+~Sc~+~Sd (panel d). The panels (e) and (f) show the subsamples of small and large Sb~+~Sc~+~Sd galaxies, selected according to the median value of the physical galaxy size shown in panel (d) of Fig.~\ref{fig:F01}. The shaded regions enclosing the profiles correspond to the standard deviations of the mean $\Delta~T(\theta)$ values for the radial bins.
    In all panels, the green shaded regions correspond to the standard deviations of 20 random realisations with the same number of centres than the galaxy data.
   }
   \label{fig:FA3}
\end{figure*}

The presence of hot diffuse gas in galactic halos can lead to the thermal Sunyaev-Zeldovich effect, which could affect CMB measures in the vicinity of galaxies. In order to explore whether this may influence our results, we have analysed temperature profiles around galaxies considering the CMB temperature SMICA map from which Sunyaev-Zeldovich sources have been projected out, as provided in the Planck database of PR3 \citep{planck2018IV}.
With this in mind,  
we have considered the same sample of galaxies analysed previously.
The results of this test are shown in Fig.~\ref{fig:FA3} which shows essentially the same 
patterns of the previous analysis performed. Therefore, we conclude that the SZ effect is not a significant driver of the results obtained in our work.
%XXX
%
%

% Don't change these lines
\bsp	% typesetting comment
\label{lastpage}
\end{document}